\documentclass[12pt]{article}
\usepackage{amsmath, amssymb}
\usepackage{mathtools}
\usepackage{booktabs}
\usepackage{longtable}
\usepackage{hyperref}
\input{definition}
\usepackage{algorithm}
\usepackage{algorithmic}

\begin{document}
\begin{center}
{\bf\large Distributed Convoluted Rank Regression for Non-Shareable Data under Non-Additive Losses}
\\\vskip0.3cm
  Wen Zhang, Liping Zhu and Songshan Yang
\\\vskip0.3cm
\textit{Renmin University of China}
\begin{singlespace}
	\footnotetext[1]{
 }
\end{singlespace}
\end{center}

\begin{singlespace}
\begin{abstract}
We study high-dimensional rank regression when data are distributed across multiple machines and the loss is a non-additive U-statistic, as in convoluted rank regression (CRR). Classical communication-efficient surrogate likelihood (CSL) methods crucially rely on the additivity of the empirical loss and therefore break down for CRR, whose global loss couples all sample pairs across machines. We propose a distributed convoluted rank regression (DCRR) framework that constructs a similar surrogate loss and demonstrate its validity under the non-additive losses. We show that this surrogate shares the same population minimizer as the full-data CRR loss and yields estimators that are statistically equivalent to centralized CRR. Building on this, we develop a two-stage sparse DCRR procedure---an iterative $\ell_1$-penalized stage followed by a folded-concave refinement---and establish non-asymptotic error bounds, a distributed strong oracle property, and a DHBIC-type criterion for consistent model selection. A scaling result shows that the number of machines may diverge as $M = o({N/(s^2\log p)})$ while achieving centralized oracle rates with only $O(\log N)$ communication rounds. Simulations and a large-scale real data example demonstrate substantial gains over naive divide-and-conquer, particularly under heavy-tailed errors.

\end{abstract}
\end{singlespace}
\newpage
\section{Introduction}

High-dimensional regression arises in many contemporary applications such as genomics, imaging, and finance, where both the sample size $N$ and the ambient dimension $p$ can be large and the data may be stored across multiple machines. In addition, the noise distribution is often heavy-tailed or contaminated by outliers, so that least-squares based methods become unreliable. Rank-based regression offers a robust alternative that is insensitive to monotone transformations and outliers in the response; early seminal works include \citet{jurevckova1971nonparametric} and \citet{jaeckel1972estimating}, and we refer to \citet{hettmansperger2010robust} for a comprehensive treatment. Recently, convoluted rank regression (CRR) has been proposed as a smooth rank-type $U$-statistic loss that retains the robustness of traditional rank estimators while enjoying differentiability and favorable efficiency \citep{zhou2024sparse}. In high dimensions, \citet{zhou2024sparse} established sparsity oracle properties for penalized CRR, and related high-dimensional robust regression methods have been studied for more general $M$-estimators; see, for example, \citet{wang2020tuning} and \citet{loh2017statistical}. 

Despite this progress, existing CRR theory and methodology are essentially centralized: they assume that all data are stored on a single machine and that the quadratic number of pairwise terms in the CRR loss can be evaluated and differentiated. 
This becomes problematic when $N$ is large and the data are distributed over multiple machines due to memory, privacy, or organizational constraints. There is a growing literature on communication-efficient distributed estimation for large-scale problems; see, for example, \citet{zhang2013communication,jordan2019communication,battey2018distributed,fan2023communication} and the references therein. Many of these works focus on least-squares or generalized linear models \citep{zhang2013divide,chen2014split,rosenblatt2016optimality,wang2017efficient,lee2017communication,huang2019distributed}, where the empirical loss is additive over observations. Distributed robust and quantile regression have also been studied under additive loss structures, including distributed quantile regression and its one-step refinements \citep{chen2020distributed,Pan02102022} and debiased Lasso based robust estimation \citep{shi2020communication}. 

A key reason for the success of these methods is that most communication-efficient surrogate likelihood (CSL) and related frameworks critically rely on an additivity assumption on the empirical loss function. Specifically, for a broad class of $M$-estimators, the full-data loss can be written as the average of local losses,
\[
\mathcal L_N(\bbeta)
  = \frac{1}{M}\sum_{m=1}^M \mathcal L_m(\bbeta),
  \qquad
\nabla \mathcal L_N(\bbeta)
  = \frac{1}{M}\sum_{m=1}^M \nabla \mathcal L_m(\bbeta),
\]
where $\mathcal L_m$ is computed using data stored on the $m$-th machine. This exact equality allows one to reconstruct the full-data gradient from local gradients and to construct a surrogate loss on a single master machine that mimics the global loss, leading to highly communication-efficient procedures \citep{zhang2013communication,jordan2019communication,fan2023communication}. On top of this additive structure, recent works have further established oracle properties for distributed procedures in high-dimensional models: \citet{gu2025collaborative} propose a collaborative score-type test with an oracle property for the test statistic in a general high-dimensional model, and \citet{liu2025communication} develop communication-efficient distributed sparse learning algorithms that enjoy oracle properties and geometric convergence for nonconvex penalized $M$-estimators. However, all these oracle results are derived under empirical risks of the form $\frac{1}{N}\sum_{i=1}^N \ell(\bbeta; z_i)$ and rely on the additivity of the global loss across machines. To the best of our knowledge, there are currently no distributed oracle guarantees for high-dimensional estimators under non-additive U-statistic rank losses, where the global empirical loss cannot be decomposed as an average of local losses. Filling this gap is one of the main goals of the present paper.

In convoluted rank regression, by contrast, the empirical loss takes the form of a second-order $U$-statistic over all sample pairs \citep{zhou2024sparse}. The global CRR loss couples observations across machines through cross-machine pairs, whereas each local loss only involves pairs within a single machine. Consequently, the full CRR loss is not the simple average of local losses, and the corresponding gradients do not satisfy
\[
\nabla \mathcal L_N(\bbeta)
  = \frac{1}{M}\sum_{m=1}^M \nabla \mathcal L_m(\bbeta).
\]
This fundamental mismatch between the $U$-statistic structure of CRR and the additivity assumption underlying CSL implies that existing CSL theory and its variants, including the distributed oracle results of \citet{gu2025collaborative} and \citet{liu2025communication}, cannot be directly applied to distributed rank regression. In particular, naive attempts to replace $\nabla \mathcal L_N(\bbeta)$ by the average of local CRR gradients introduce an additional bias.
and it is far from obvious whether one can still obtain centralized optimal rates and strong oracle properties for sparse CRR estimators in high dimensions.

In this paper, we develop a distributed sparse rank regression framework that overcomes this difficulty and focuses on high-dimensional estimation, variable selection, and model selection. Our key observation is that, although the global CRR loss and the local CRR losses are no longer additively related at the sample level, they share the same population risk function due to the common kernel and the i.i.d.\ sampling scheme. In particular, the population limits of the global and local CRR losses coincide, and hence they possess the same minimizer $\bbeta^\star$. We leverage this fact to construct a novel surrogate loss that only requires the CRR $U$-statistic loss on a single ``master'' machine together with a gradient correction term aggregated from all machines. This surrogate loss no longer enjoys exact equality with the full-data loss, but we show that it has the same population minimizer and that, under suitable conditions, its minimizer is statistically equivalent to the centralized CRR estimator in high-dimensional regimes. Technically, this requires new tools to control the discrepancy between the true full-data gradient of the $U$-statistic loss and our gradient approximation based on local $U$-statistics, using concentration inequalities for sub-Gaussian processes and 
$U$-statistics 
\citep{bousquet2003concentration,ledoux2013probability},
and to establish restricted curvature properties for the surrogate loss in sparse directions.

Building on this surrogate, we propose a two-stage distributed convoluted rank regression (DCRR) procedure for high-dimensional sparse estimation. In the first stage, we compute an $\ell_1$-penalized DCRR estimator via an iterative distributed algorithm that communicates only gradients of local CRR losses. In the second stage, we refine this initial estimator by applying a folded-concave penalty through a local linear approximation scheme, in the spirit of nonconvex penalization \citep{fan2001variable,zhang2010nearly} and the I-LAMM framework \citep{Fan2015ILAMMFS}. This yields a computationally scalable estimator that combines the robustness of rank regression, the efficiency of CRR, and the statistical advantages of folded-concave regularization in high dimensions. We further develop a distributed high-dimensional BIC (DHBIC) criterion for tuning parameter selection and show that it consistently recovers the true support of $\bbeta^\star$ in a distributed environment.

Compared with existing high-dimensional CRR methods \citep{zhou2024sparse}, our framework scales to massive distributed data sets without requiring computation of the full quadratic CRR loss. Compared with distributed regression procedures under additive losses \citep{zhang2013communication,battey2018distributed,fan2023communication,chen2020distributed,Pan02102022,gu2025collaborative,liu2025communication}, our method is specifically tailored to non-additive $U$-statistic rank losses and, to the best of our knowledge, provides the first communication-efficient distributed solution with strong oracle guarantees for high-dimensional rank regression under a non-additive loss.

\subsection{Our contributions}

We summarize our main contributions as follows.

\textbf{A communication-efficient framework for $U$-statistic rank regression adapted to non-additive losses.
}
Since the global CRR loss involves cross-machine pairs and is no longer the average of local losses \citep{zhou2024sparse}, this introduces new challenges to our analysis.
To remedy this, we exploit the fact that the global and local CRR losses share the same population risk and minimizer, and we construct the surrogate loss that combines the CRR loss on a single machine with a gradient correction term computed from local CRR gradients.
We prove that this surrogate loss has the same population minimizer as the full-data CRR loss, and that its minimizer in high dimensions attains the same convergence rate as the centralized CRR estimator under suitable sparsity and regularity conditions. This extends the CSL philosophy \citep{zhang2013communication,jordan2019communication,fan2023communication} to a broad class of $U$-statistic based rank regression problems where exact sample-level additivity fails.

\textbf{Distributed strong oracle property for high-dimensional rank regression.}
  Based on the surrogate loss, we develop a two-stage distributed CRR estimator that first applies $\ell_1$-penalization and then refines the solution via folded-concave penalties \citep{fan2001variable,zhang2010nearly} using a local linear approximation scheme \citep{Fan2015ILAMMFS}. We establish non-asymptotic error bounds and show that, with an appropriate number of communication rounds, the folded-concave DCRR estimator enjoys a \emph{distributed strong oracle property}: it recovers the true support of $\bbeta^\star$ with high probability and achieves the same estimation rate as the ideal oracle estimator that knows the support in advance, matching the centralized high-dimensional CRR benchmark \citep{zhou2024sparse}. In contrast to recent distributed oracle results for additive losses \citep{gu2025collaborative,liu2025communication}, our oracle property is obtained under a non-additive $U$-statistic rank loss where the global empirical loss is not the sum of local losses. This requires new bias control and curvature arguments that are specific to the distributed CRR setting. Moreover, our framework allows the number of machines $M$ to diverge with the total sample size $N$, subject to mild high-dimensional scaling such as $M = o\{N/(s^2 \log p)\}$, and centralized oracle rates and model selection consistency can be achieved with only $O(\log N)$ communication rounds. This is substantially more permissive than naive divide-and-conquer schemes, whose statistical accuracy is fundamentally constrained by the local sample size. Finally, our algorithm remains tuning-free for the bandwidth parameter $h$, which just need to satisfy $h=O(1)$ and does not require data-dependent selection. 

\textbf{Distributed model selection for rank regression.}
  We introduce a distributed HBIC-type criterion (DHBIC) for tuning parameter selection in high-dimensional rank regression and prove its model selection consistency. The DHBIC selector yields sparse models that recover the true support with high probability under mild conditions, while avoiding sample splitting or repeated refitting. Combined with our estimation results, this provides a fully distributed procedure for sparse high-dimensional CRR that is both statistically accurate and computationally efficient.

\medskip
The rest of the paper is organized as follows. In Section~\ref{section:2}, we introduce the distributed CRR framework, define the DCRR surrogate loss, and present the $\ell_1$-penalized and folded-concave penalized algorithms. Section~\ref{section:3} establishes non-asymptotic error bounds, the distributed strong oracle property, and the consistency of the DHBIC model selection criterion. Section~\ref{section:4} reports simulation studies and a real-data application that illustrate the numerical performance of the proposed methods. We conclude with a brief discussion in Section~\ref{sec:discussion}.

\subsection{Notation}
We introduce some necessary notations. 
For $a,b\in\mR$, $a\wedge b=\min\{a,b\}$ and $a\vee b=\max\{a,b\}$. For a matrix $A$, $\lambda_{\min}(A)$ and $\lambda_{\max}(A)$ represent the smallest and the largest absolute value of eigenvalues.
For a vector $\a$, $\supp(\a)=\{j\mid a_j\neq 0\}$. For a positive integer $p$, $[p]=\{1,2,\ldots,p\}$.
For an index set $\calA$, a vector $\b$ and matrix $\C$, $\b_{\calA}=(b_j,j\in\calA)$ and $\C_{\calA\calA}=(c_{ij},i\in\calA ,j\in\calA)$. 
For a vector $\a=(a_1,\ldots,a_p)$ and $q\in[1,\infty)$, define the $l_q$ norm as $\norm{\a}_q=(\sum_{j=1}^p a_j^q)^{\frac{1}{q}}$. Let $\norm{\a}_{\infty}=\max_{j}\abs{a_j}$ be the $l_{\infty}$ norm and $\norm{\a}_{\min}=\min_{j}\abs{a_j}$.
For a symmetric positive-definite matrix $\B$, define the vector norm as $\norm{\a}_{\B}=\norm{\B^{1/2}\a}_2$.
For a sequence $a_n$ and another nonnegative sequence $\{b_n\}$,
we write $a_n=O(b_n)$ when there exist a constant $C>0$ such that $\abs{a_n}\leq Cb_n$; We use $a_n=o(b_n)$ when $a_n/b_n\rightarrow0$. We write $a_n\lesssim b_n$ when there exist a constant $C>0$ such that $a_n\leq Cb_n$; we write $a_n\asymp b_n$ if $a_n\lesssim b_n$ and $b_n\lesssim a_n$. 
For two random sequences $Z_n$ and $Z_n'$, we write $Z_n=O_p(Z_n')$ when for any $\epsilon>0$, there exists a constant $M>0$ and $N$, $\P(\abs{Z_n/Z_n'}>M)<\epsilon$ for all $n\geq N$; we write $Z_n=o_p(1)$ when $\lim_{n\rightarrow\infty}\P(\abs{Z_n/Z_n'}>\epsilon)=0$, for any $\epsilon>0$.

\section{Distributed Convoluted Rank Regression}
\label{section:2}

In this section, we introduce the distributed convoluted rank regression (DCRR) framework. We first review the canonical convoluted rank regression (CRR) and then describe our distributed surrogate loss and sparse high-dimensional estimators.

\subsection{Model and convoluted rank regression}

We consider a distributed setting where the data are stored on $M$ machines. For notational convenience, we present our methodology and theory under a balanced design in which each machine holds the same number $n$ of observations so that the total sample size satisfies $N = nM$; see Remark~\ref{rem:unequal-n} for a discussion of heterogeneous local sample sizes. Suppose we observe identically distributed data $\{(y_{i}^{m},\x_{i}^{m}): i=1,\ldots,n,\; m=1,\ldots,M\}$,
where $\{(y_i^{m},\x_i^{m})\}_{i=1}^n$ denotes the subsample stored on the $m$-th machine. Here $y_{i}^m\in \mR$ is the response and $\x_{i}^m\in\mR^p$ is the $p$-dimensional covariate vector. Let $\X=(\X_1,\ldots,\X_p)\in \mR^{N\times p}$ be the global design matrix and $\X^m=(\X_1^m,\ldots,\X_p^m)\in \mR^{n\times p}$ be the design matrix on the $m$-th machine, where $\X_j=(x_{1j},\ldots,x_{Nj})\trans$ and $\X_j^m=(x_{1j}^m,\ldots,x_{nj}^m)\trans$ contain observations of the $j$-th variable, $j=1,\ldots,p$. We use the same notation for the response vectors $\y$ and $\y^m$.

Assume that the data are generated from the linear regression model
\begin{equation}
\label{equation:1}
y_i^m=\x_i^m\bbeta^\ast+\epsilon_i^m,
\quad i=1,\ldots,n,\; m=1,\ldots,M,
\end{equation}
where $\{\epsilon_i^m: i=1,\ldots, n, m=1,\ldots,M\}$ are i.i.d.\ random errors, and $\bbeta^\ast\in \mR^p$ is the unknown parameter vector. Without loss of generality, we assume that the errors in \eqref{equation:1} have mean zero and are independent of $\x$. As clarified in Remark~\ref{rem:unequal-n}, all our results extend to the more general case with unequal local sample sizes $n_m$ by choosing any machine with sufficiently large $n_m$ as the master machine.

Let $(y',\x')$ be an independent copy of $(y,\x)$. For rank regression, the true regression coefficient $\bbeta^\ast$ can be characterized as the minimizer of the population rank loss
\[
\bbeta^\ast=\arg \min_{\bbeta\in\mR^p}\mE\big\{\abs{y-y'-(\x-\x')\trans\bbeta}\big\}. 
\]
However, the non-smooth absolute loss brings substantial challenges for computation and high-dimensional analysis. To address this, \citet{zhou2024sparse} proposed the canonical convoluted rank regression
\begin{equation}
\label{equation:2}
\min_{\bbeta\in \mR^p} \frac{1}{N(N-1)}\sum_{i=1}^{N}\sum_{j\neq i}
L_h\big\{y_i-y_j-(\x_i-\x_j)\trans\bbeta\big\},
\end{equation}
where
$L_h(u)=\int_{-\infty}^{\infty}\abs{u-v}\frac{1}{h}K\!\left(\frac{v}{h}\right)\,dv$
is a smooth convex function obtained as the convolution of $L(u)=\abs{u}$ and $K_h(u)=\frac{1}{h}K(u/h)$. 

We assume that the kernel $K$ satisfies the following properties:  
(i) $K(-t)=K(t)$ for all $t \in \mathbb{R}$;  
(ii) there exists $\delta_0>0$ such that $\kappa_l:=\inf _{t \in[-\delta_0, \delta_0]} K(t)>0$;  
(iii) $\int_{-\infty}^{\infty} K(t) \mathrm{d} t=1$;  
(iv) $\kappa_u:=\sup _{t \in \mathbb{R}} K(t)<\infty$;  
(v) $\kappa_j:=\int_{-\infty}^{\infty}|t|^j K(t) \mathrm{d} t<\infty$ for $j=1,2$;  
(vi) there exist $\alpha_0 \in(0,1]$ and $L_0>0$ such that 
$|K(x)-K(y)| \leq L_0|x-y|^{\alpha_0}$ for all $x, y \in \mathbb{R}$.  
Typical examples include the Gaussian kernel $K(u)=\frac{1}{\sqrt{2\pi}}\exp\{-u^2/2\}$ and the Epanechnikov kernel $K(u)=\frac{3}{4}(1-u^2)\I(-1\leq u \leq 1)$.

Let $\mathcal L_h(\bbeta)=\mE\big\{L_h\big(y-y'-(\x-\x')\trans\bbeta\big)\big\}$ and denote its minimizer by $\bbeta_h^\ast$. \citet{zhou2024sparse} showed that for any $h>0$, $\bbeta_h^\ast=\bbeta^\ast$,
so the smoothing does not introduce asymptotic bias at the population level.

\subsection{Distributed CRR and surrogate loss}

In a distributed system, the full CRR loss in \eqref{equation:2} is not directly available, because it involves all $N(N-1)$ pairs across machines. We define the global and local empirical losses as
\begin{align*}
\calL_N(\bbeta)
&=\frac{1}{N(N-1)}\sum_{i=1}^{N}\sum_{j\neq i}
L_h\big(y_i-y_j-(\x_i-\x_j)\trans\bbeta\big),\\
\calL_m(\bbeta)
&=\frac{1}{n(n-1)}\sum_{i=1}^{n}\sum_{j\neq i}
L_h\big(y_i^m-y_j^m-(\x_i^m-\x_j^m)\trans\bbeta\big),
\quad m=1,\ldots,M.
\end{align*}
The global loss $\calL_N$ is infeasible to compute in a distributed environment when $N$ is large, whereas each $\calL_m$ is computable on the $m$-th local machine.

For general $M$-estimators with an additive empirical loss of the form
\[
\calL_N(\bbeta)=\frac{1}{M}\sum_{m=1}^M \calL_m(\bbeta).
\]
\citet{jordan2019communication} proposed the communication-efficient surrogate likelihood (CSL) framework. In its simplest form, CSL constructs the surrogate loss
\begin{equation}
\label{equation:3}
\wt{\calL}(\bbeta)=\calL_1(\bbeta)
-\Inner{\bbeta}{\nabla\calL_1(\bbeta_0)-\nabla\calL_N(\bbeta_0)},
\end{equation}
where $\bbeta_0$ is an initial estimator for $\bbeta^\ast$ and the first machine is used as the center. This construction crucially relies on the additivity $\nabla\calL_N(\bbeta)=M^{-1}\sum_{m=1}^M\nabla\calL_m(\bbeta)$, which allows the global gradient to be reconstructed from local gradients.

In convoluted rank regression, however, the empirical loss is a second-order $U$-statistic over all sample pairs. The global CRR loss $\calL_N$ couples observations across machines through cross-machine pairs, whereas each $\calL_m$ only involves within-machine pairs. As a consequence,
\[
\frac{1}{M}\sum_{m=1}^M\nabla\calL_m(\bbeta)\neq \nabla \calL_N(\bbeta),
\]
so the additivity assumption underlying CSL fails. This makes existing CSL theory inapplicable to distributed CRR and raises the question of whether we can still construct a communication-efficient surrogate that achieves the same statistical performance as a centralized CRR estimator.

Inspired by the form of \eqref{equation:2} and the CSL idea in \eqref{equation:3}, we propose the distributed convoluted rank regression (DCRR) estimator
\begin{equation}
\label{equation:dcr}
\wh\bbeta_d=\arg\min_{\bbeta\in\mR^p}
\Big\{
\calL_1(\bbeta)
-\Inner{\bbeta}{\nabla\calL_1(\bbeta_0)-\frac{1}{M}\sum_{m=1}^M\nabla\calL_m(\bbeta_0)}
\Big\},
\end{equation}
which uses only the CRR loss on the central machine and a gradient correction aggregated from local machines. Define the distributed convoluted rank loss
\[
\calL_d(\bbeta;\bbeta_0)
:=\calL_1(\bbeta)
-\Inner{\bbeta}{\nabla\calL_1(\bbeta_0)-\frac{1}{M}\sum_{m=1}^M\nabla\calL_m(\bbeta_0)}.
\]
By construction, $\calL_d(\cdot;\bbeta_0)$ is not equal to the infeasible full-data loss $\calL_N$, and $\frac{1}{M}\sum_{m=1}^M\nabla\calL_m(\bbeta)$ does not coincide with $\nabla\calL_N(\bbeta)$ due to the $U$-statistic structure. Nevertheless, we will show that they share the same population minimizer and that the resulting estimator enjoys the same limiting distribution and oracle rate as centralized CRR.

Let $\bbeta_d^\ast=\arg\min_{\bbeta\in\mR^p}\mE\{\calL_d(\bbeta;\bbeta_0)\}$. The following lemma shows that the population minimizer of our surrogate loss coincides with the true parameter $\bbeta^\ast$.

\begin{lemm}
\label{lemm:1}
For any $h>0$ and any $M\geq1$, we have $\bbeta_d^\ast=\bbeta^\ast$.
\end{lemm}

This conclusion follows directly from the equality $\bbeta_h^\ast = \bbeta^\ast$ for the population CRR loss, and it provides a population-level justification for our surrogate construction.

We next consider the asymptotic distribution of the DCRR estimator. The following lemma shows that, under a mild condition on the initial estimator, $\wh\bbeta_d$ is asymptotically equivalent to the centralized CRR estimator.

\begin{lemm}
\label{lemm:2}
For each $h>0$, if $\norm{\bbeta_0-\bbeta^\ast}_2=o_p(M^{-1/2})$, then
\[
\sqrt{N}(\wh\bbeta_d-\bbeta^\ast)
\xrightarrow{d}\calN\left(
\bzeros,\,
\frac{\mE\big[\mE\{L_h'(\epsilon-\epsilon')\mid\epsilon\}^2\big]}{\mE\{L_h''(\epsilon-\epsilon')\}^2}
\cov(\x)^{-1}
\right),
\]
where $(\epsilon,\epsilon')$ is an independent copy of the error pair and $\cov(\x)=\Sigma$ is the covariance matrix of $\x$.
\end{lemm}

\begin{rema}
Lemma~\ref{lemm:2} shows that, under the initial condition $\norm{\bbeta_0-\bbeta^\ast}_2=o_p(M^{-1/2})$, the DCRR estimator $\wh\bbeta_d$ has the same asymptotic variance as the global convoluted estimator $\wh\bbeta_h=\arg\min_{\bbeta}\calL_N(\bbeta)$. Let $\wh\bbeta=\arg\min_{\bbeta}\frac{1}{N(N-1)}\sum_{i=1}^N\sum_{j\neq i}\abs{y_i-y_j-(\x_i-\x_j)\trans\bbeta}$ be the nonsmoothed rank estimator. Then the asymptotic relative efficiency satisfies
\[
\ARE(\wh\bbeta_d,\wh\bbeta)
=\frac{\mE\big[\mE\{L_h'(\epsilon-\epsilon')\mid\epsilon\}^2\big]}{12\left\{\int f^2(x)\,dx\right\}^2\mE\{L_h''(\epsilon-\epsilon')\}^2},
\]
where $f(\cdot)$ is the density of $\epsilon$. From Theorem~3 of \citet{zhou2024sparse}, we have $\lim_{h\rightarrow0^+}\ARE(\wh\bbeta_d,\wh\bbeta)=1$, which means that when $h$ is chosen sufficiently small, the estimation efficiency of $\wh\bbeta_d$ can be made arbitrarily close to that of $\wh\bbeta$. Together with the high-dimensional results in Section~\ref{section:3}, this shows that DCRR achieves centralized CRR efficiency while only using local $U$-statistics and one round of communication.
\end{rema}

\subsection{Sparse DCRR and iterative algorithms}

In high-dimensional settings, the dimension $p$ can be large and may grow with $N$, while we often assume that only a relatively small subset of coordinates of $\bbeta^\ast$ have non-negligible effects on the outcome. Let $\calA=\{j: \beta_j^\ast\neq 0\}$ be the support of $\bbeta^\ast$ and $s=\abs{\calA}\geq1$ be its size. We allow both $p$ and $s$ to diverge with $n$, and assume that $s$ is of smaller order than $n$.

To incorporate sparsity, we consider the penalized DCRR estimator
\begin{equation}
\label{pen-dcrr}
\min_{\bbeta\in \mR^p} \left\{
\calL_d(\bbeta;\bbeta_0)
+\sum_{j=1}^{p}p_{\lambda}(\abs{\beta_j})
\right\},
\end{equation}
where $p_{\lambda}(\cdot)$ is a penalty function with tuning parameter $\lambda$, such as the $\ell_1$ penalty or a folded-concave penalty. 
We now describe a two-stage iterative algorithm that performs $\ell_1$-penalized iterations in the first stage, followed by a second-stage refinement based on folded-concave penalties. Theoretical properties of these estimators are presented in Section~\ref{section:3}.

In the first stage, we first perform an initial iteration using the DCRR with an $l_1$ penalty.
We design the $\ell_1$-penalized DCRR iterative algorithm as
\begin{equation}
\label{eq:l1_stage}
{\wh{\bbeta}}^{k}:=\arg\min_{\bbeta\in\mR^p}
\Big\{
\calL_1(\bbeta)
-\Inner{\bbeta}{\nabla\calL_1({\wh{\bbeta}}^{k-1})-\frac{1}{M}\sum_{m=1}^M\nabla\calL_m({\wh{\bbeta}}^{k-1})}
+\lambda\sum_{j=1}^{p}\abs{\beta_j}
\Big\},
\end{equation}
for $k=1,\ldots,k_1$, where $k_1\geq1$ is the maximum number of iterations. At iteration $k$, we use the previous estimate $\wh{\bbeta}^{k-1}$ as the initial value in the DCRR loss. The center machine broadcasts $\wh{\bbeta}^{k-1}$ to all local machines, each local machine computes the gradient $\nabla\calL_m(\wh{\bbeta}^{k-1})$, and the center aggregates these gradients to form the surrogate loss. Each iteration incurs communication cost of order $O(pM)$, so the $\ell_1$-penalized DCRR algorithm is communication-efficient.

Motivated by the strong oracle properties of folded-concave penalized estimators \citep{fan2001variable,zhang2010nearly}, we employ a second-stage refinement based on folded-concave penalties to enhance estimation accuracy and achieve distributed strong oracle guarantees. 

For the folded-concave penalized DCRR, we adopt a local linear approximation (LLA) to $p_{\lambda}(\cdot)$ and consider penalties of the form $p_\lambda(v)=\lambda^2 q(v / \lambda)$ for $v \geq 0$, where $p:[0, \infty) \rightarrow[0, \infty)$ satisfies:  
(i) $p(\cdot)$ is non-decreasing and concave on $[0, \infty)$ with $p(0)=0$;  
(ii) $p^{\prime}(\cdot)$ exists almost everywhere, is non-increasing on $(0, \infty)$, $0 \leq p^{\prime}(v) \leq 1$, and $\lim _{v \downarrow 0} q^{\prime}(v)=1$;  
(iii) $p^{\prime}(\alpha_1)=0$ for some $\alpha_1>0$.  
Special cases include the SCAD penalty \citep{fan2001variable} and the MCP penalty \citep{zhang2010nearly}.

For convenience, we denote the $k$-th iterate of the $\ell_1$-penalized DCRR algorithm by $\wh\bbeta^{(1,k)}$. Let the initial estimator for the second stage be $\wh{\bbeta}^{(2,0)}=\wh{\bbeta}^{(1,k_1)}$, and let $k_2$ be the maximum number of iterations in the second stage. Using the LLA idea, for $t\geq 2$ and $k=1,\ldots,k_t$, we define the $t$-th stage estimator $\wh{\bbeta}^{(t,k)}$ as
\begin{equation}
\label{eq:multi_stage}
\wh{\bbeta}^{(t,k)}=\arg\min_{\bbeta\in\mR^p}
\Big\{
\calL_d(\bbeta;\wh\bbeta^{(t,k-1)})
+\sum_{j=1}^{p} p^{\prime}_{\lambda_{t,k}}\big(\abs{\wh{\beta}_j^{(t,0)}}\big)\abs{\beta_{j}}
\Big\},
\end{equation}
where $\wh{\bbeta}^{(t,0)}=\wh\bbeta^{(t-1,k_t)}$. In practice, we will mainly focus on the case $k_t=1$ for $t\geq2$, which already suffices to achieve the distributed strong oracle property described in Section~\ref{section:3}. The corresponding simplified second-stage algorithm is summarized in Algorithm~\ref{algo:2}.

\begin{algorithm}[H]
    \caption{Two-Stage DCRR Iterative Algorithm}
    \label{algo:2}
    \begin{algorithmic}[1]
    \STATE \textbf{Input:} bandwidth $h$, maximum number of iterations $k_1$, maximum number of stages $T$, initial value 
    $\wh{\bbeta}^0=\arg\min_{\bbeta\in \mR^p}
    \Big\{\calL_1(\bbeta)+\lambda\sum_{j=1}^{p}\abs{\beta_j}\Big\}$.
     \FOR{$k=1,\ldots, k_1$}
        \STATE \textbf{Center:} transmit the current iterate $\wh{\bbeta}^{(1,k-1)}$ to all local machines;
        \STATE \textbf{Local:} each machine $m$ computes the local gradient $\nabla \calL_m(\wh{\bbeta}^{(1,k-1)})$ and sends it to the center;
        \STATE \textbf{Center:} update the estimator by \eqref{eq:l1_stage}.
    \ENDFOR

    \FOR{$t=2,\ldots, T$}
        \STATE \textbf{Center:} set $\wh\bbeta^{(t,0)}=\wh\bbeta^{(t-1)}$;
        \STATE \textbf{Center:} transmit the current iterate $\wh{\bbeta}^{(t,0)}$ to all local machines;
        \STATE \textbf{Local:} each machine $m$ computes the local gradient $\nabla \calL_m(\wh{\bbeta}^{(t,0)})$ and sends it to the center;
        \STATE \textbf{Center:} update the estimator by
        \[
        {\wh{\bbeta}}^{(t)}:=\arg\min_{\bbeta\in\mR^p}
        \Big\{
        \calL_1(\bbeta)
        -\Inner{\bbeta}{\nabla\calL_1({\wh{\bbeta}}^{(t,0)})-\frac{1}{M}\sum_{m=1}^M\nabla\calL_m({\wh{\bbeta}}^{(t,0)})}
        +\sum_{j=1}^{p}p_{\lambda}'\big(\abs{\beta_j}\big)\abs{\beta_j}
        \Big\}.
        \]
    \ENDFOR
    \RETURN $\wh\bbeta^{(T)}$.
    \end{algorithmic}
\end{algorithm}


\section{Theoretical Properties of Sparse DCRR}
\label{section:3}

In this section, we establish non-asymptotic error bounds and oracle properties for the proposed sparse DCRR estimators. We first introduce regularity assumptions and notation, 
then study the convergence of the two-stage DCRR iterative algorithm,
the distributed oracle property, and tuning parameter selection via distributed HBIC.

\subsection{Assumptions and notation}

We begin with regularity conditions on the error distribution and the design.

\begin{assum}[Error distribution]
\label{assum:1}
Let $g(\cdot)$ be the probability density function of $\zeta_{ij}=\epsilon_i-\epsilon_j$. We assume that $g$ is Lipschitz continuous: there exists a constant $L_{1}>0$ such that 
$|g(x)-g(y)| \leq L_{1}|x-y|$ for all $x, y \in \mathbb{R}$. 
This implies $\mu_{0}:=\sup _{t \in \mathbb{R}} g(t)<\infty$. Moreover, we assume that there exist positive constants $\delta_{1}, \mu_{1}$ such that 
$g(t) \geq \mu_{1}$ for all $t \in[-\delta_{1}, \delta_{1}]$.
\end{assum}

\begin{assum}[Design]
\label{assum:2}
The covariate vector $\x$ has bounded components and zero mean:
$\norm{\x}_{\infty} \leq b_1$ almost surely and $\mE(\x)=\bzeros$. Let $\x'$ be an independent copy of $\x$ and define $\z=\Sigma^{-1/2}(\x-\x')$, where $\Sigma=\mE(\x\x\trans)$. We assume that $\z$ is sub-Gaussian, i.e., there exists a constant $v_1\geq 1$ such that
$\P\big(\abs{\z\trans\z'}>v_1\norm{\z'}_2 t\big)\leq 2e^{-t^2/2}$
for all $\z'\in\mR^p$ and $t\geq 0$.
\end{assum}

\begin{assum}[Restricted eigenvalue condition]
\label{assum:3}
Let $\Sigma=\mE(\x\x\trans)$. There exists a constant $\rho>0$ such that 
$\min_{\v\in\mR^p}\frac{\norm{\v}_{\Sigma}^2}{\norm{\v}_2^2}\geq \rho$,
where $\norm{\v}_{\Sigma}^2=\v\trans\Sigma\v$. In particular, $\lambda_{\min}(\Sigma)\geq \rho$ and $\lambda_{\max}(\Sigma)<\infty$.
\end{assum}

Assumption~\ref{assum:1} is a Lipschitz condition on the density of the error difference, which is slightly stronger than the H\"older condition in \citet{zhou2024sparse} but simplifies our exposition. Assumption~\ref{assum:2} is standard for random design, and we implicitly work with centered covariates so that no intercept is needed. And the sub-Gaussian condition is used to derive high-dimensional concentration bounds. Assumption~\ref{assum:3} is a restricted eigenvalue type condition that allows us to control the $\ell_2$ norm via the $\Sigma$-norm through $\norm{\v}_2\leq \rho^{-1/2}\norm{\v}_{\Sigma}$. In fact, we only require this to hold on certain cone sets introduced below. 

For $d,l>0$, we define the local elliptical region and $\ell_1$ cone
\[
\Theta(d)=\{\v\in \mR^p:\, \norm{\v}_{\Sigma}\leq d\},
\qquad
\Lambda(l)=\{\v\in\mR^p:\, \norm{\v}_1\leq l\norm{\v}_{\Sigma}\}.
\]

We also introduce shrinkage factors for the 
folded-concave penalized DCRR estimators. 
\begin{equation}
\label{def:gamma}
\gamma =\wt C_3\Big(\sqrt{\frac{s+x}{nh}}+\sqrt{\frac{s+x}{Nh}}\Big),\qquad \lambda^\ast= C_2\sqrt{\frac{\log p}{N}},
\end{equation}
where $\wt C_3=24C_3v_1^2\kappa^{-1}\rho^{-1/2}$ and $C_3=258e\kappa_u\mu_0$, and $x>0$ is a tuning parameter appearing in concentration bounds; $ C_2=4\sqrt{2}(1+c_0^2)^{1/2}b_1$ for some $c_0>0$. We also assume that there exists $\alpha_0>0$ such that
\[
\phi=\frac{\big[1+\{p'(\alpha_0)/2\}^2\big]^{1/2}}{\alpha_0\kappa\rho^{1/2}}\in(0,1),
\]
and define
\[
l=\Big\{2+\frac{2}{p'(\alpha_0)}\Big\}
\Big(\frac{c^2+1}{\rho}\Big)^{1/2}s^{1/2},
\]
where $c>0$ satisfies $p'(\alpha_0)(c^2+1)^{1/2}/2+1=\alpha_0\kappa\rho^{1/2}c$. These constants will be used to characterize the contraction behavior of the iterative algorithms.

\begin{rema}[Different local sample sizes]
\label{rem:unequal-n}

More generally, suppose the $m$-th machine contains $n_m$ observations with $N = \sum_{m=1}^M n_m$, and let $\calL_m$ be the corresponding local CRR loss based on $n_m$ samples. If we choose as the master machine any index $m_\star$ with local sample size $n_\star = n_{m_\star}$, then the DCRR surrogate loss and algorithms are defined exactly as before with $\calL_1$ replaced by $\calL_{m_\star}$. All arguments in Section~\ref{section:3} carry over with $n$ replaced by $n_\star$ in the conditions of Theorems~\ref{theo:2}--\ref{theo3}. In particular, the key requirement is that $n_\star$ satisfies the same high-dimensional scaling as $n$, for example $s^2 \log p / n_\star = o(1)$ and $M = o\{N/(s^2 \log p)\}$. Since $\max_{1\le m\le M} n_m \ge N/M$, we can always select the machine with the largest local sample size as the master, ensuring that these conditions are met whenever the global scaling assumptions hold. 
\end{rema}

\subsection{Convergence of the Two-Stage DCRR Iterative Algorithm}
We now study the non-asymptotic error bounds of the two-stage DCRR iterative algorithm. 
For simplicity, let 
\[
r_{t,k}=\norm{\wh\bbeta^{(t,k)}-\bbeta^\ast}_{\Sigma},
\qquad
r_{t,0}=\norm{\wh\bbeta^{(t,0)}-\bbeta^\ast}_{\Sigma},
\]
and define the index set
$\calT_{t,k}
=\calA\cup\big\{j\in[p]:\lambda_j^{(t,k)}<p'(\alpha_0)\lambda_{t,k}\big\}$,
where $\mathbf{\lambda}^{(t,k)}=(\lambda_1^{(t,k)},\ldots,\lambda_p^{(t,k)})\trans$ with $\lambda_j^{(t,k)}=p^{\prime}_{\lambda_{t,k}}(\abs{\wh{\beta}_j^{(t,0)}})$. We assume that $\abs{\calT_{t,k}}\leq (c^2+1)s$, which is a sparsity condition on the initial estimator and is naturally satisfied by the $\ell_1$-penalized estimator.
Let $\bome^\ast=\frac{1}{M}\sum_{m=1}^M\nabla\calL_m(\bbeta^{\ast})$ and denote its restriction to $\calA$ by $\bome^\ast_{\calA}$.

\begin{lemm}
\label{theo:2}
Assume Assumptions~\ref{assum:1}--\ref{assum:3} and the condition $\abs{\calT_{t,k}}\leq (c^2+1)s$ hold. Let $h\gtrsim\{s\log p/n+\sqrt{x/n}\}$ and $\lambda_{t,k}\asymp\gamma_{1} r_{t,k-1} +\lambda^\ast$. Then for any $t\geq 2$ and $1\leq k\leq k_t$, the second-stage estimator $\wh{\bbeta}^{(t,k)}$ satisfies: $\wh\bbeta^{(t,k)}\in\Lambda(l)$, $\supp(\wh\bbeta^{(t,k)})\lesssim s$; and
    \begin{equation}
    \label{theo:2equ}
    r_{t,k}
    \leq \gamma r_{t,k-1}+\phi r_{t,0}
    +\kappa^{-1}\rho^{-1/2}
    \Big\{
    \big\|\bome^\ast_{\calA}\big\|_2
    +
    \big\|p_{\lambda_{t,k}}^{\prime}\big((\abs{\bbeta_{\calA}^\ast}-\alpha_0\lambda_{t,k})_{+}\big)\big\|_2
    \Big\}.
    \end{equation}
with probability at least $1-10e^{-x}$.
\end{lemm}

\begin{rema}
From \eqref{theo:2equ}, we obtain
\[
r_{t,k} 
\leq \{\gamma^k+\phi(1-\gamma^k)/(1-\gamma)\}r_{t,0}
+\frac{1-\gamma^k}{1-\gamma}\,
\kappa^{-1}\rho^{-1/2}
\Big\{
\big\|\bome^\ast_{\calA}\big\|_2
+\big\|p_{\lambda_{t,k}}^{\prime}\big((\abs{\bbeta_{\calA}^\ast}-\alpha_0\lambda_{t,k})_{+}\big)\big\|_2
\Big\}.
\]
This shows that the key to achieving further shrinkage beyond the $\ell_1$-penalized estimator is the existence of $k\geq1$ such that $\gamma^k+\phi(1-\gamma^k)/(1-\gamma)<1$, which is equivalent to $\gamma+\phi<1$. When $s\log p/n=o(1)$, taking $x=c_0\log p$ and $h=O(1)$ makes $\gamma+\phi<1$ natural. Thus, under the scaling $s\log p/n=o(1)$, our algorithm is  tuning-free with respect to the bandwidth $h$. 
The three additional terms in \eqref{theo:2equ} reflect the initial error $r_{t,0}$, the oracle error $\norm{\bome^\ast_{\calA}}_2$, and the bias introduced by the penalty. As proved in Lemma~B5 of the Supplementary Material, $\norm{\bome^\ast_{\calA}}_2=O_p\big(\{(s+x)/N\}^{1/2}\big)$.
\end{rema}

To simplify the second stage, we focus on the case $k_t=1$ for $t\geq2$. In this case, write $\wh\bbeta^{(t)}$ for $\wh\bbeta^{(t,1)}$ and choose $\lambda=\lambda_{t,1}\asymp(\log p/N)^{1/2}$. 

\begin{theo}
\label{theo3}
Assume Assumptions~\ref{assum:1}--\ref{assum:3}, the beta-min condition $\norm{\bbeta_{\calA}^\ast}_{\min}>(\alpha_0+\alpha_1)\lambda$, $x'\lesssim s\log p$, and $s^2\log p/n=o(1)$ hold. 
Let $h=O(1)$ and $\lambda\asymp(\log p/N)^{1/2}$. Then, for the $(t+1)$-th 
estimator $\wh{\bbeta}^{(t+1)}$,
\[
\norm{\wh{\bbeta}^{(t+1)}-\bbeta^\ast}_{\Sigma}\lesssim \sqrt{\frac{s+x'}{N}},
\]
with probability at least $1-10t p^{-c_0}-e^{-x'}$, provided that 
$
t\gtrsim \frac{\log\{\log p+\log(c_0+1)\}}{\log(1/\bar{\gamma})}$,
 where $\bar{\gamma}=\gamma+\phi$.
\end{theo}

\begin{rema}
Theorem~\ref{theo3} provides an oracle rate for the final two-stage estimator. 
In fact, we can prove that after sufficiently many $\ell_1$ iterations as $k_1=O_p\{\log(1/(s^{1/2}\lambda^\ast))\}$, the first-stage estimator provides a near-oracle estimator with rate $(s\log p/N)^{1/2}$ and satisfies $\left\{\wh\bbeta^{(1,k_1)}-\bbeta^\ast\in\Lambda(l),\;\supp(\wh\bbeta^{(1,k_1)})\lesssim s\right\}$.  Further details of the proofs are provided in the Supplementary Materials.
Then, the folded-concave second stage improves the error to the oracle rate $(s/N)^{1/2}$, up to logarithmic factors absorbed in $x'$. There is no additional bias due to smoothing thanks to Lemma~\ref{lemm:1}. 
Under the beta-min condition, the term $\big\|p_{\lambda_{t,k}}^{\prime}\big((\abs{\bbeta_{\calA}^\ast}-\alpha_0\lambda_{t,k})_{+}\big)\big\|_2$ vanishes.
\end{rema}
\subsection{Distributed oracle property}

Folded-concave penalized estimators are known to enjoy strong oracle properties in centralized settings. In our distributed setting, however, the full-sample oracle estimator is infeasible because the global CRR loss $\calL_N$ is not available. 
It is therefore challenging to directly impose a strong oracle property relative to the full-sample folded-concave CRR estimator. 
Instead, we introduce a \emph{distributed oracle estimator} defined in terms of the surrogate loss $\calL_d$ and study the distance between our practical estimator and this distributed oracle.

For $t\geq 2$, we define the $t$-th distributed oracle estimator as
\begin{equation}
\label{def: ora}
\wh{\bbeta}^{\ora,t}=\underset{\bbeta_{\calA^c}=0}{\arg \min }
\Big\{
\calL_1(\bbeta)
-\Inner{\bbeta}{\nabla\calL_1(\wh\bbeta^{(t-1)})-\frac{1}{M}\sum_{m=1}^M\nabla\calL_m(\wh\bbeta^{(t-1)})}
\Big\}.
\end{equation}

\begin{theo}
\label{theo4}
Under the conditions of Theorem~\ref{theo3}, the $(t+1)$-th distributed oracle estimator satisfies
\[
\norm{\wh{\bbeta}^{\ora,t+1}-\bbeta^\ast}_{\Sigma}\lesssim \sqrt{\frac{s+x'}{N}},
\]
with probability at least $1-10t p^{-c_0}-e^{-x'}$, provided that $t\gtrsim\frac{\log\{\log p+\log(c_0+1)\}}{\log(1/\gamma)}$.
\end{theo}

Theorem~\ref{theo4} shows that the distributed oracle estimator achieves the same oracle rate $O_p\{(s/N)^{1/2}\}$ as the centralized oracle estimator based on the full data. This oracle benchmark is nontrivial in our setting because the global CRR loss is non-additive and cannot be decomposed as a sum of local losses. In the next theorem, we show that our practical folded-concave DCRR estimator coincides with the distributed oracle estimator with high probability, thus yielding a distributed strong oracle property.

We impose an additional irrepresentable-type condition on the covariance structure.

\begin{assum}[Irrepresentable condition]
\label{assum:5}
There exists a constant $A_0\geq 0$ such that 
$
\max_{j\in\calA^c}\norm{\Sigma_{j\calA}(\Sigma_{\calA\calA})^{-1}}_1\leq A_0,
$
where $\Sigma_{j\calA}=\mE(\x_j\x_{\calA}\trans)$ and $\Sigma_{\calA\calA}=\mE(\x_{\calA}\x_{\calA}\trans)$.
\end{assum}

\begin{rema}[On Assumption~\ref{assum:5}]
Classical irrepresentable conditions for the Lasso typically require
$
\max_{j\in\calA^c}
\big\|
\Sigma_{j\calA}(\Sigma_{\calA\calA})^{-1}
\big\|_1
\le 1-\eta$ for some $\eta>0$,
that is, the constant is strictly smaller than $1$. In contrast,
Assumption~\ref{assum:5} only asks that this quantity be bounded by a
finite constant $A_0\ge 0$, and our analysis does not use the stronger
requirement $A_0<1$.
\end{rema}

\begin{theo}
\label{theo5}
Assume Assumptions~\ref{assum:1}--\ref{assum:5} and the beta-min condition $\norm{\bbeta_{\calA}^\ast}_{\min}\geq (\alpha_0+\alpha_1)\lambda$ hold, 
$n\gtrsim \max\{N/(s^3\log p),\, N/\{s(s+\log p)\}\}$ and $s^2\log p/n=o(1)$. Let $h=O(1)$ and $\lambda\asymp(\log p/N)^{1/2}$. Then the distributed strong oracle property $\wh{\bbeta}^{t+1}=\wh{\bbeta}^{\ora,t+1}$
holds with probability at least $1-12t p^{-c_0}-p^{-1}$, provided that 
$t\gtrsim \log\big(\abs{\calA^{(1)}}^{1/2}\big)$, where $\calA^{(1)}=\supp(\wh\bbeta^{(2,0)})\cup\calA$.
\end{theo}

\begin{rema}
Theorem~\ref{theo5} proves that, after a logarithmic number of folded-concave refinement steps, the second-stage DCRR estimator coincides with the distributed oracle estimator with high probability. Combined with Theorem~\ref{theo4}, this yields a distributed strong oracle property for high-dimensional DCRR 
This is qualitatively different from existing distributed oracle results in the literature, which rely on additive empirical losses where the global objective is the sum of local objectives. In our setting, the global CRR loss is not additive across machines, yet the proposed DCRR procedure still attains the same oracle rate and model selection performance as a centralized oracle estimator. 

According to the conditions of Theorems~\ref{theo:2} and~\ref{theo3}, the number of machines $M$ can diverge at a reasonable speed as: 
$M = o\Big(\frac{N}{s^2 \log p}\Big)$ and $
M \lesssim \max\{s^3\log p,s(s+\log p)\}$.
Choose $k_1$ and $T$ so that the contraction requirements in Theorems~\ref{theo:2} and~\ref{theo3} are satisfied, and let $R = k_1 + T - 1$ denote the total number of gradient communication rounds.
In particular, DCRR attains the centralized high-dimensional CRR oracle rate and exact support recovery under a diverging number of machines $M$, while requiring only $R = O\big(\log N + \log\log p\big)$ gradient communication rounds between the master and local machines. 
In other words, as long as each machine has enough observations, 
the proposed distributed estimator behaves, both in estimation error and support recovery, as if we had centralized access to all $N$ samples, while using only a logarithmic number of communication rounds.
\end{rema}

\subsection{Tuning parameter selection: distributed HBIC}

In practice, a data-driven approach is needed to select tuning parameters. The HBIC proposed by \citet{zhou2024sparse} is not directly applicable in a distributed system, because it requires access to the full-data CRR loss. We therefore propose a distributed high-dimensional BIC (DHBIC) criterion tailored to our DCRR framework.

For $t\geq 2$, define
\[
\DHBIC(\lambda)=
\log\left\{
\frac{1}{M}\sum_{m=1}^{M}\calL_m(\wh{\bbeta}_{\lambda}^{(t)})
\right\}
+\abs{\supp(\wh{\bbeta}_{\lambda}^{(t)})}\frac{C_N\log p}{n},
\]
where 
$\wh{\bbeta}_{\lambda}^{(t)}
=\arg\min_{\bbeta\in\mR^p}
\Big\{
\calL_d(\bbeta;\wh\bbeta^{(t-1)})
+\sum_{j=1}^p p_{\lambda}(\abs{\beta_j})
\Big\}$.
We choose the tuning parameter by $\wh\lambda=\arg\min_{\lambda\in \calM^{(t)}}\DHBIC(\lambda)$,
where $\calM^{(t)}=\{\lambda>0:\, \abs{\supp(\wh\bbeta_{\lambda}^{(t)})}\leq K_N\}$ and $K_N>s$ is allowed to diverge.

\begin{theo}
\label{theo:dhbic}
Under the conditions of Theorem~\ref{theo5}, suppose $\mE\abs{\epsilon-\epsilon'}<\infty$, $K_N=o\{(N/\log p)\wedge C_N\}$, $C_Ns\log p/N=o(1)$, and 
\[
\Big(\frac{C_Ns^{1/2}\log p}{N}\Big)^{1/2}
\vee
\Big\{\frac{C_Ns^{1/2}K_N\log p}{N}\Big\}
=o\big(\norm{\bbeta_{\calA}^\ast}_{\min}\big).
\]
Then the DHBIC selector satisfies $\supp(\wh\bbeta_{\wh\lambda}^{(t+1)})=\calA$
with probability at least $1-12t p^{-c_0}-C_NN^{-1/2}-4p^{-1}$, provided that 
$t\gtrsim\log\big(\abs{\calA^{(1)}}^{1/2}\big)$.
\end{theo}

\begin{rema}
Theorem~\ref{theo:dhbic} guarantees that the tuning parameter chosen by DHBIC yields a folded-concave DCRR estimator that consistently recovers the true support $\calA$, thereby preserving the distributed strong oracle property of Theorems~\ref{theo5}.
Similar to HBIC in centralized settings, DHBIC does not require sample splitting or repeated refitting and therefore enjoys low computational complexity. In practice, one can choose $C_N\asymp\log(\log N)$.
\end{rema}

\section{Numerical Studies}
\label{section:4}

In this section, we assess the finite-sample performance of the proposed distributed convoluted rank regression (DCRR) procedures. The simulations are designed to illustrate (i) the estimation accuracy and variable selection performance of DCRR relative to centralized CRR, 
(ii) the gain from the folded-concave refinement and its connection to the distributed strong oracle property, (iii) the robustness of rank-based methods under heavy-tailed errors, and (iv) the limitations of naive divide-and-conquer strategies. We then apply the methods to a real-data example.

\subsection{Simulation design}

We consider the linear model $y=\x\trans\bbeta^\ast+\epsilon$,
where $\bbeta^\ast=(\sqrt{3},\sqrt{3},\sqrt{3},0,\dots,0)\trans\in \mR^p$ so that the true support size is $s=3$. The covariate vector $\x$ is generated from a mean-zero multivariate normal distribution $\calN(0,\Sigma)$ with an autoregressive correlation structure
$\Sigma_{ij}=0.5^{\abs{i-j}}$, $1\leq i,j\leq p$,
denoted by AR(0.5). We set the ambient dimension to $p=1000$. The total sample size is $N=nM$ with local sample size $n=100$ on each machine and consider $M=5$ and $M=15$ machines, corresponding to $N=500$ and $N=1500$, respectively.

To examine robustness to heavy-tailed errors, we generate the error term $\epsilon$ from three distributions of increasing tail heaviness:(1) standard normal: $\epsilon\sim \calN(0,1)$;
(2) $t$-distribution with 4 degrees of freedom, scaled to unit variance: $\epsilon\sim \sqrt{2}\,t(4)$;
(3) standard Cauchy distribution: $\epsilon\sim \mathrm{Cauchy}(0,1)$.
Thus, the design progressively moves from well-behaved Gaussian noise to moderately heavy-tailed and extremely heavy-tailed settings.

Throughout the simulations, we use the Epanechnikov kernel $K(u)=\frac{3}{4}(1-u^2)\I(-1\leq u\leq1)$ in the CRR loss and fix the bandwidth at $h=1$, which is compatible with the theoretical conditions in Section~\ref{section:3}. For each combination of $(\epsilon,M)$, we generate $100$ independent replicates.

We compare the following estimators:
\begin{table}[htbp]
\centering
\scalebox{0.8}{
\begin{tabular}{ll}
\toprule
\textbf{Category} & \textbf{Estimator Description} \\
\midrule

Centralized CRR &
\begin{tabular}[t]{@{}l@{}}
CRR-LASSO: $\ell_1$-penalized CRR using the full sample; \\
CRR-SCAD: SCAD-penalized CRR using the full sample.
\end{tabular} \\[4pt]

Distributed DCRR (proposed) &
\begin{tabular}[t]{@{}l@{}}
DCRR-LASSO: $\ell_1$-penalized first-stage estimator; \\
DCRR-SCAD ($T=2$): folded-concave DCRR with $T=2$ refinements; \\
DCRR-SCAD ($T=6$): folded-concave DCRR with $T=6$ refinements.
\end{tabular} \\[4pt]

Divide-and-conquer CRR (baseline) &
\begin{tabular}[t]{@{}l@{}}
DC-CRR-LASSO: one-shot divide-and-conquer CRR-LASSO; \\
DC-CRR-SCAD: analogous estimator with SCAD penalty.
\end{tabular} \\[4pt]

Oracle methods &
\begin{tabular}[t]{@{}l@{}}
CRR-ORA: centralized oracle CRR estimator; \\
DCRR-ORA ($T=2$) / ($T=6$): oracle DCRR with restricted support $\calA$.
\end{tabular} \\
\bottomrule
\end{tabular}
}
\end{table}

To evaluate estimation and variable selection performance, we compute: $\ell_1$ estimation error: $\mE\norm{\wh{\bbeta}-\bbeta^\ast}_1$; $\ell_2$ estimation error: $\mE\norm{\wh{\bbeta}-\bbeta^\ast}_2$; FP: the average number of false positives (selected noise variables); FN: the average number of false negatives (missed true signals).
All performance measures are averaged over the $100$ replicates, and we report the means with Monte Carlo standard errors (SEs) in parentheses. For the CRR-based methods, tuning parameters are selected by HBIC \citep{zhou2024sparse} (centralized CRR) or the proposed DHBIC in Section~\ref{section:3} (DCRR). For DC-CRR methods, HBIC is applied independently on each local machine. 

\subsection{Simulation results}

Tables~\ref{tab:normal}--\ref{tab:cauchy} summarize the results under $\epsilon\sim \calN(0,1)$, $\epsilon\sim \sqrt{2}\,t(4)$, and $\epsilon\sim\mathrm{Cauchy}(0,1)$, respectively. Within each scenario, the best performance with respect to each criterion is highlighted in bold.

\begin{table}[htbp]
\centering
\caption{Simulation results under an AR(0.5) design and $\epsilon \sim \calN(0,1)$ for $M=5$ and $M=15$. We compare centralized CRR, distributed CRR (DCRR), and divide-and-conquer CRR (DC-CRR), each with either an $\ell_1$ or SCAD penalty, along with oracle estimators for CRR and DCRR. Reported are mean (SE) over 100 replicates.}
\label{tab:normal}
\scalebox{0.8}{
\begin{tabular}{llcccc}
\toprule
$M$ & Method & $\ell_1$ & $\ell_2$ & FP & FN \\
\midrule
\multirow{12}{*}{$M=5$} 
 & CRR-LASSO       & 0.34(0.01) & 0.22(0.01) & 0.01(0.01) & 0.00(0.00) \\
 & CRR-SCAD        & 0.14(0.01) & 0.09(0.01) & \textbf{0.00}(0.00) & \textbf{0.00}(0.00) \\
 & DCRR-LASSO      & 0.32(0.01) & 0.21(0.01) & 0.03(0.02) & 0.00(0.00) \\
 & DCRR-SCAD (T=2) & 0.15(0.01) & 0.10(0.01) & \textbf{0.00}(0.00) & \textbf{0.00}(0.00) \\
 & DCRR-ORA (T=2)  & 0.15(0.01) & 0.10(0.01) & 0.00(0.00) & 0.00(0.00) \\
 & CRR-ORA         & \textbf{0.13}(0.01) & \textbf{0.08}(0.00) & 0.00(0.00) & 0.00(0.00) \\
 & DCRR-SCAD (T=6) & \textbf{0.13}(0.01) & 0.09(0.00) & \textbf{0.00}(0.00) & \textbf{0.00}(0.00) \\
 & DCRR-ORA (T=6)  & \textbf{0.13}(0.01) & 0.09(0.00) & 0.00(0.00) & 0.00(0.00) \\
 & DC-CRR-LASSO    & 0.81(0.01)&0.50(0.01)&0.78(0.08)&0.00(0.00)\\
 & DC-CRR-SCAD     & 0.33(0.02)&0.21(0.01)&0.82(0.08)&0.00(0.00)\\
\midrule
\multirow{12}{*}{$M=15$} 
 & CRR-LASSO       & 0.20(0.01)&0.13(0.00)&0.03(0.02)&0.00(0.00)\\
 & CRR-SCAD        & 0.09(0.00)&0.06(0.00)&\textbf{0.00}(0.00)&\textbf{0.00}(0.00)\\
 & DCRR-LASSO      & 0.17(0.00)&0.11(0.00)&0.00(0.00)&0.00(0.00)\\
 & DCRR-SCAD (T=2) & 0.09(0.00)&0.06(0.00)&\textbf{0.00}(0.00)&\textbf{0.00}(0.00)\\
 & DCRR-ORA (T=2)  & 0.09(0.00)&0.06(0.00)&0.00(0.00)&0.00(0.00)\\
 & CRR-ORA         & \textbf{0.08}(0.00)&\textbf{0.05}(0.00)&0.00(0.00)&0.00(0.00)\\
 & DCRR-SCAD (T=6) & \textbf{0.08}(0.00)&\textbf{0.05}(0.00)&\textbf{0.00}(0.00)&\textbf{0.00}(0.00)\\
 & DCRR-ORA (T=6)  & \textbf{0.08}(0.00)&\textbf{0.05}(0.00)&0.00(0.00)&0.00(0.00)\\
 & DC-CRR-LASSO    &0.79(0.01)&0.48(0.00)&2.38(0.14)&0.00(0.00)\\
 & DC-CRR-SCAD     &0.30(0.01)&0.19(0.01)&2.47(0.17)&0.00(0.00)\\
\bottomrule
\end{tabular}
}
\end{table}

\begin{table}[htbp]
\centering
\caption{Simulation results under an AR(0.5) design and $\epsilon \sim \sqrt{2}\,t(4)$ for $M=5$ and $M=15$. Mean (SE) over 100 replicates.}
\label{tab:t4}
\scalebox{0.8}{
\begin{tabular}{llcccc}
\toprule
$M$ & Method & $\ell_1$ & $\ell_2$ & FP & FN \\
\midrule
\multirow{12}{*}{$M=5$} 
 & CRR-LASSO        & 0.56(0.02) & 0.37(0.01) & 0.06(0.02) & 0.00(0.00) \\
 & CRR-SCAD         & 0.25(0.01) & 0.16(0.01) & 0.00(0.00) & 0.00(0.00) \\
 & DCRR-LASSO       & 0.55(0.01) & 0.35(0.01) & 0.02(0.01) & 0.00(0.00) \\
 & DCRR-SCAD (T=2)  & 0.26(0.01) & 0.17(0.01) & 0.00(0.00) & 0.00(0.00) \\
 & DCRR-ORA (T=2)   & 0.25(0.01) & 0.17(0.01) & 0.00(0.00) & 0.00(0.00) \\
 & CRR-ORA          & \textbf{0.23}(0.01) & \textbf{0.15}(0.01) & 0.00(0.00) & 0.00(0.00) \\
 & DCRR-SCAD (T=6)  & \textbf{0.23}(0.01) & \textbf{0.15}(0.01) & 0.00(0.00) & 0.00(0.00) \\
 & DCRR-ORA (T=6)   & \textbf{0.23}(0.01) & \textbf{0.15}(0.01) & 0.00(0.00) & 0.00(0.00) \\
 & DC-CRR-LASSO     & 1.40(0.02) & 0.85(0.01) & 0.98(0.10) & 0.00(0.00) \\
 & DC-CRR-SCAD      & 1.00(0.04) & 0.64(0.03) & 1.51(0.11) & 0.00(0.00) \\
\midrule
\multirow{12}{*}{$M=15$}
 & CRR-LASSO        & 0.35(0.01) & 0.23(0.00) & 0.12(0.03) & 0.00(0.00) \\
 & CRR-SCAD         & 0.15(0.01) & 0.10(0.00) & \textbf{0.00}(0.00) & \textbf{0.00}(0.00) \\
 & DCRR-LASSO       & 0.31(0.01) & 0.20(0.00) & 0.01(0.01) & 0.00(0.00) \\
 & DCRR-SCAD (T=2)  & 0.15(0.01) & 0.10(0.00) & \textbf{0.00}(0.00) & \textbf{0.00}(0.00) \\
 & DCRR-ORA (T=2)   & 0.15(0.01) & 0.10(0.00) & 0.00(0.00) & 0.00(0.00) \\
 & CRR-ORA          & \textbf{0.11}(0.01) & \textbf{0.08}(0.00) & 0.00(0.00) & 0.00(0.00) \\
 & DCRR-SCAD (T=6)  & 0.12(0.01) & \textbf{0.08}(0.00) & \textbf{0.00}(0.00) & \textbf{0.00}(0.00) \\
 & DCRR-ORA (T=6)   & 0.12(0.01) & \textbf{0.08}(0.00) & 0.00(0.00) & 0.00(0.00) \\
 & DC-CRR-LASSO     & 1.43(0.01) & 0.86(0.01) & 2.72(0.14) & 0.00(0.00) \\
 & DC-CRR-SCAD      & 0.94(0.02) & 0.60(0.01) & 3.38(0.18) & 0.00(0.00) \\
\bottomrule
\end{tabular}
}
\end{table}

\begin{table}[htbp]
\centering
\caption{Simulation results under an AR(0.5) design and $\epsilon \sim \mathrm{Cauchy}(0,1)$ for $M=5$ and $M=15$. Mean (SE) over 100 replicates.}
\label{tab:cauchy}
\scalebox{0.8}{
\begin{tabular}{llcccc}
\toprule
$M$ & Method & $\ell_1$ & $\ell_2$ & FP & FN \\
\midrule
\multirow{12}{*}{$M=5$}
 & CRR-LASSO        & 0.86(0.09) & 0.54(0.05) & 0.01(0.01) & 0.10(0.05) \\
 & CRR-SCAD         & 0.47(0.10) & 0.29(0.06) & 0.00(0.00) & 0.10(0.05) \\
 & DCRR-LASSO       & 0.79(0.08) & 0.50(0.05) & 0.00(0.00) & 0.09(0.05) \\
 & DCRR-SCAD (T=2)  & 0.50(0.09) & 0.32(0.05) & 0.00(0.00) & 0.09(0.05) \\
 & DCRR-ORA (T=2)   & 0.31(0.02) & 0.21(0.01) & 0.00(0.00) & 0.00(0.00) \\
 & CRR-ORA          & \textbf{0.24}(0.01) & \textbf{0.16}(0.01) & 0.00(0.00) & 0.00(0.00) \\
 & DCRR-SCAD (T=6)  & 0.44(0.09) & 0.28(0.05) & 0.00(0.00) & 0.09(0.05) \\
 & DCRR-ORA (T=6)   & 0.26(0.01) & 0.17(0.01) & 0.00(0.00) & 0.00(0.00) \\
 & DC-CRR-LASSO     & 3.73(0.07) & 2.16(0.04) & 0.08(0.03) & 0.07(0.03) \\
 & DC-CRR-SCAD      & 2.79(0.08) & 1.71(0.04) & 0.11(0.03) & 0.02(0.01) \\
\midrule
\multirow{12}{*}{$M=15$}
 & CRR-LASSO        & 0.48(0.05) & 0.31(0.03) & 0.01(0.01) & 0.02(0.02) \\
 & CRR-SCAD         & 0.23(0.05) & 0.15(0.03) & 0.00(0.00) & 0.02(0.02) \\
 & DCRR-LASSO       & 0.40(0.04) & 0.26(0.02) & 0.00(0.00) & 0.01(0.01) \\
 & DCRR-SCAD (T=2)  & 0.22(0.04) & 0.14(0.02) & 0.00(0.00) & 0.01(0.01) \\
 & DCRR-ORA (T=2)   & 0.18(0.01) & 0.11(0.01) & 0.00(0.00) & 0.00(0.00) \\
 & CRR-ORA          & 0.14(0.01) & 0.09(0.00) & 0.00(0.00) & 0.00(0.00) \\
 & DCRR-SCAD (T=6)  & \textbf{0.19}(0.04) & \textbf{0.12}(0.03) & 0.00(0.00) & 0.01(0.01) \\
 & DCRR-ORA (T=6)   & 0.15(0.01) & 0.10(0.00) & 0.00(0.00) & 0.00(0.00) \\
 & DC-CRR-LASSO     & 3.80(0.04) & 2.20(0.02) & 0.26(0.04) & 0.00(0.00) \\
 & DC-CRR-SCAD      & 2.89(0.05) & 1.75(0.03) & 0.36(0.06) & 0.00(0.00) \\
\bottomrule
\end{tabular}
}
\end{table}

\noindent\textbf{DCRR versus centralized CRR.}
Under Gaussian errors (Table~\ref{tab:normal}), the DCRR-LASSO estimator closely tracks CRR-LASSO in both $\ell_1$ and $\ell_2$ errors for $M=5$ and $M=15$, confirming that the surrogate loss preserves the statistical efficiency of centralized CRR in a distributed environment. 
In practice, we have set the maximum number of iterations for the DCRR-LASSO $k_1=8$.  
The folded-concave DCRR-SCAD estimators with $T=2$ and $T=6$ further reduce the estimation error and essentially match the centralized CRR-SCAD in both estimation error and variable selection (FP and FN are essentially zero). 
This is consistent with the oracle rates established in Theorems~\ref{theo3}. 


\noindent\textbf{Effect of the folded-concave refinement and oracle benchmarks.}
Across Tables~\ref{tab:normal}–\ref{tab:cauchy}, the oracle CRR estimator (CRR-ORA) provides a lower bound on the achievable error. The oracle DCRR estimators DCRR-ORA($T=2$) and DCRR-ORA($T=6$) are numerically indistinguishable from CRR-ORA, demonstrating that the distributed surrogate loss does not degrade the oracle performance. For non-oracle estimators, a single folded-concave refinement ($T=2$) already brings DCRR-SCAD very close to CRR-SCAD, and additional refinements ($T=6$) yield further, albeit modest, improvements. This mirrors the theoretical picture: the first stage achieves a near-oracle rate, while the second stage improves towards the oracle rate.

\noindent\textbf{Naive divide-and-conquer CRR.}
The DC-CRR-LASSO and DC-CRR-SCAD estimators perform substantially worse than both centralized CRR and DCRR in all settings, especially for larger $M$ and heavier-tailed errors. Their $\ell_2$ errors are several times larger than those of CRR-SCAD and DCRR-SCAD, and they exhibit non-negligible FP even when FN is zero. This confirms that naive averaging of local CRR estimators cannot effectively correct the local estimation bias, and the resulting accuracy is limited by the local sample size $n$, in sharp contrast to the proposed DCRR procedure, which achieves full-sample rates under the same distributed architecture.


\noindent\textbf{Summary.}
Overall, the simulations demonstrate that:
(i) DCRR-LASSO and DCRR-SCAD achieve estimation and selection performance comparable to their centralized CRR counterparts while operating in a communication-efficient distributed manner;  
(ii) the folded-concave refinement improves the first-stage estimator towards the oracle benchmark, in line with the distributed strong oracle property in Theorem~\ref{theo5};  
(iii) naive divide-and-conquer CRR can be substantially suboptimal;  
(iv) the DCRR procedures are robust to heavy-tailed errors.

\subsection{Real-data application}

Used-car price modeling provides a canonical example where heavy-tailed response distributions and massive, naturally distributed data arise. Vehicle prices can vary by orders of magnitude, with a small fraction of high-end or luxury vehicles producing extreme outliers, and transaction records are typically collected and stored across different dealers, online platforms, and insurers. This makes used-car pricing a natural test bed for robust, communication-efficient distributed regression methods. 

We illustrate the proposed methods on a real data set of used cars in the United States from Kaggle (\url{https://www.kaggle.com/datasets/ananaymital/us-used-cars-dataset}). After data cleaning, we retain $N_{full} = 332{,}382$ records corresponding to sport utility vehicles (SUVs). The response is the selling price of a vehicle, measured in units of \$1,000.
We consider 27 base predictors, including both continuous and categorical variables such as horsepower, mileage, model year, engine size, and manufacturer. A detailed description of these variables can be found in Figure~1 of \citet{Pan02102022}.

To construct a high-dimensional design, we create dummy variables for all categorical predictors and include two-way interactions among selected predictors. After removing highly collinear columns to ensure numerical stability, we obtain $p = 604$ predictors (excluding the intercept). The price of the vehicle is right-skewed and displays signs of heavy tails, which makes it a challenging setting for least-squares methods and a natural testbed for rank-based approaches.

To make centralized CRR methods computationally feasible and to mimic repeated sampling from a large population, we proceed as follows. In each replicate, we randomly draw a subsample of size $N=2000$ without replacement from the full data. The first $N_{\mathrm{train}}=1000$ observations are used as a training set, and the remaining $N_{\mathrm{test}}=1000$ as a test set. For distributed methods, the $N_{\mathrm{train}}$ observations are evenly partitioned across $M\in\{25,20,10,8,5\}$ machines. Within each replicate, we fit:
\begin{itemize}
    \item[1.] distributed methods: DCRR-LASSO, DCRR-SCAD($T=2$), DCRR-SCAD($T=6$), DC-CRR-LASSO, and DC-CRR-SCAD;
    \item[2.] centralized methods: CRR-LASSO, CRR-SCAD.
\end{itemize}
For each fitted model, we compute the $\ell_1$ and $\ell_2$ prediction errors on the test set and record the model size (MS), defined as the number of nonzero coefficients. All results are averaged over 100 replications. The tuning parameters are selected as in the simulation study. Table~\ref{tab:realdata} reports the results.

\begin{table}[htbp]
\centering
\caption{Real-data results for the used cars data set. For each number of machines $M$, we report the average $\ell_1$ and $\ell_2$ prediction errors on the test set and the model size (MS). Results are averaged over 100 random subsamples; SEs are in parentheses.}
\label{tab:realdata}
\scalebox{0.8}{
\begin{tabular}{llccc}
\toprule
$M$ & Method & $\ell_1$ & $\ell_2$ & MS \\
\midrule
\multirow{5}{*}{$M=25$}
& DCRR-LASSO        & 4.64(0.04)          & 7.26(0.13)          & 11.13(0.36)   \\
& DCRR-SCAD(T=2)    & 4.28(0.03)          & 6.61(0.13)          & 4.92(0.22)    \\
& DCRR-SCAD(T=6)    & \textbf{4.18(0.03)} & \textbf{6.52(0.13)} & \textbf{3.30(0.10)} \\
& DC-CRR-LASSO      & 7.76(0.03)          & 10.97(0.11)         & 50.89(0.96)   \\
& DC-CRR-SCAD       & 7.75(0.03)          & 10.97(0.11)         & 51.20(0.91)   \\
\midrule
\multirow{5}{*}{$M=20$}
& DCRR-LASSO        & 4.38(0.04)          & 6.96(0.14)          & 8.73(0.27)    \\
& DCRR-SCAD(T=2)    & 4.19(0.03)          & 6.52(0.13)          & 4.59(0.16)    \\
& DCRR-SCAD(T=6)    & \textbf{4.10(0.03)} & \textbf{6.44(0.13)} & \textbf{3.76(0.13)}    \\
& DC-CRR-LASSO      & 7.73(0.03)          & 10.94(0.11)         & 40.49(0.56)   \\
& DC-CRR-SCAD       & 7.71(0.03)          & 10.92(0.12)         & 41.77(0.59)   \\
\midrule
\multirow{5}{*}{$M=10$}
& DCRR-LASSO        & 4.29(0.02)          & 6.87(0.13)          & 7.43(0.18)    \\
& DCRR-SCAD(T=2)    & 4.06(0.02)          & 6.44(0.12)          & 4.71(0.21)    \\
& DCRR-SCAD(T=6)    & \textbf{4.03(0.02)} & \textbf{6.39(0.12)} & \textbf{4.20(0.13)}    \\
& DC-CRR-LASSO      & 7.47(0.03)          & 10.60(0.12)         & 5.39(0.20)    \\
& DC-CRR-SCAD       & 5.81(0.04)          & 8.61(0.13)          & 16.20(0.36)   \\
\midrule
\multirow{5}{*}{$M=8$}
& DCRR-LASSO        & 4.23(0.02)          & 6.80(0.13)          & 7.87(0.16)    \\
& DCRR-SCAD(T=2)    & 4.01(0.02)          & 6.37(0.13)          & 4.82(0.14)    \\
& DCRR-SCAD(T=6)    & \textbf{3.96(0.02)} & \textbf{6.30(0.13)} & \textbf{4.82(0.12)}    \\
& DC-CRR-LASSO      & 6.84(0.05)          & 9.81(0.13)          & 5.95(0.20)    \\
& DC-CRR-SCAD       & 4.64(0.05)          & 7.26(0.14)          & 14.58(0.37)   \\
\midrule
\multirow{5}{*}{$M=5$}
& DCRR-LASSO        & 4.19(0.02)          & 6.76(0.12)          & 8.56(0.24)    \\
& DCRR-SCAD(T=2)    & 3.96(0.02)          & 6.33(0.13)          & 5.46(0.16)    \\
& DCRR-SCAD(T=6)    & \textbf{3.96(0.02)} &\textbf{ 6.31(0.13)} & \textbf{4.79(0.11)}    \\
& DC-CRR-LASSO      & 5.15(0.04)          & 7.87(0.14)          & 9.24(0.12)    \\
& DC-CRR-SCAD       & 4.16(0.02)          & 6.59(0.13)          & 6.82(0.21)    \\
\midrule
\multirow{5}{*}{Global}
& CRR-LASSO         & 4.18(0.02)          & 6.75(0.13)          & 9.54(0.19)    \\
& CRR-SCAD          & 3.95(0.02)          & 6.31(0.13)          & 6.28(0.13)    \\
& NULL MODEL        & 7.78(0.03)          & 11.00(0.11)         & --            \\
\bottomrule
\end{tabular}
}
\end{table}

The real-data results highlight several points. First, across all choices of $M$, DCRR-SCAD($T=2$) and DCRR-SCAD($T=6$) consistently outperform DCRR-LASSO and the divide-and-conquer baselines in terms of prediction error, often with smaller or comparable model sizes. As $M$ decreases from 25 to 5 (i.e., as each local sample size increases), the prediction errors of DCRR-SCAD steadily decrease and approach those of the centralized CRR-SCAD, illustrating the scalability and statistical efficiency of DCRR in practice.

Second, the naive DC-CRR estimators exhibit substantially larger prediction errors than DCRR for all $M$, and either very large or very unstable model sizes. This again confirms that simple averaging of local CRR estimators is not effective in this setting.


Overall, the real-data example corroborates our theoretical and simulation findings: the proposed DCRR framework attains near-centralized performance for high-dimensional rank regression in distributed environments, improves substantially over naive divide-and-conquer methods, and provides a practically appealing compromise between robustness, predictive accuracy, and model sparsity.

\section{Discussion}
\label{sec:discussion}

We have proposed a communication-efficient distributed convoluted rank regression framework for high-dimensional sparse estimation under a non-additive $U$-statistic loss. By constructing a surrogate loss that combines a single-machine CRR loss with an aggregated gradient correction, we showed that the resulting DCRR estimators achieve the same convergence rates as centralized CRR while operating in a distributed environment. Building on this surrogate, we developed an iterative two-stage DCRR algorithm, established a distributed strong oracle property, and introduced a DHBIC criterion for consistent model selection. Numerical studies with both simulated and real data demonstrated that DCRR can match or closely approximate the performance of centralized CRR, substantially outperform naive divide-and-conquer baselines and maintaining robustness under heavy-tailed errors.

Several directions merit further investigation. First, while our theoretical analysis and simulations focused on estimation, variable selection, and model selection, our framework naturally accommodates low-dimensional inference based on partially penalized DCRR estimators. Developing a systematic distributed inference theory, including confidence intervals and hypothesis tests with finite-sample calibration, is an interesting avenue for future work. Second, it would be valuable to extend the present methodology to more general pairwise loss functions and other $U$-statistic based models where the empirical loss is inherently non-additive. Finally, exploring privacy-preserving or communication-constrained variants of DCRR, and studying their trade-offs between statistical efficiency and resource usage, would further enhance the practical applicability of distributed rank regression in large-scale applications.

\begin{center}
\bibliography{reference}
\end{center}

\end{document}